\documentclass[conference]{IEEEtran}
\IEEEoverridecommandlockouts
\usepackage{cite}
\usepackage{amsmath,amssymb,amsfonts}
\usepackage{algorithmic}
\usepackage{graphicx}
\usepackage{textcomp}
\usepackage{xcolor}
\def\BibTeX{{\rm B\kern-.05em{\sc i\kern-.025em b}\kern-.08em
    T\kern-.1667em\lower.7ex\hbox{E}\kern-.125emX}}
\usepackage{listings}
\begin{document}

\title{
Evaluation of EAP Usage for Authenticating Eduroam Users in 5G Networks
\thanks{This research was partially funded by Coordenação de Aperfeiçoamento de Pessoal de Nível Superior – Brasil (CAPES) and by Rede Nacional de Pesquisa (RNP) via its 2023 Programa de Gestão de Identidades (PGId).
}
}

\author{\IEEEauthorblockN{Leonardo Azalim de Oliveira}
\IEEEauthorblockA{\textit{Computer Science Department} \\
\textit{Federal University of Juiz de Fora}\\
Juiz de Fora, Brazil \\
leonardo.azalim@ice.ufjf.br}
\and
\IEEEauthorblockN{Edelberto Franco Silva}
\IEEEauthorblockA{\textit{Computer Science Department} \\
\textit{Federal University of Juiz de Fora}\\
Juiz de Fora, Brazil \\
edelberto@ice.ufjf.br}
}

\maketitle

\begin{abstract}
The fifth generation of the telecommunication networks (5G) established the service-oriented paradigm on the mobile networks. In this new context, the 5G Core component has become extremely flexible so, in addition to serving mobile networks, it can also be used to connect devices from the so-called non-3GPP networks, which contains technologies such as WiFi. The implementation of this connectivity requires specific protocols to ensure authentication and reliability. Given these characteristics and the possibility of convergence, it is necessary to carefully choose the encryption algorithms and authentication methods used by non-3GPP user equipment. In light of the above, this paper highlights key findings resulting from an analysis on the subject conducted through a test environment which could be used in the context of the Eduroam federation.
\end{abstract}

\begin{IEEEkeywords}
5G, authentication, EAP, Eduroam
\end{IEEEkeywords}

\section{Introduction}

The fifth generation of mobile networks (5G) marked a paradigm shift introducing the so-called service-oriented (or service-based) paradigm as one of its main changes compared to previous generations.
This new approach allowed the virtualization of a significant portion of the 5G network infrastructure, thereby greatly increasing the flexibility in configuring a 5G mobile network environment. The new generation has opened possibilities that came with some challenges, such as the integration of WiFi networks with the 5G infrastructure, a situation that has been gaining attention from the Eduroam federation community.

Education Roaming, commonly abbreviated as Eduroam, is a global initiative with the primary goal of providing secure wireless internet access to members of the academic community in participating institutions. This federation allows students, researchers, and staff from Higher Education Institutions (HEIs) to use their login credentials, managed by their home institutions, to access wireless networks at any other participating institution.

The Brazilian Eduroam federation is managed by Rede Nacional de Pesquisa (RNP)\footnote{https://www.rnp.br/servicos/eduroam}, a national institution of research that also provides network-related services. The Brazilian Eduroam already supports Hotspot 2.0 and Passpoint technologies. If these protocols are fully implemented, additional roaming options could be incorporated through the technologies already adopted by the service. These characteristics contribute to the growing interest of RNP in becoming a member of the OpenRoaming\footnote{https://wballiance.com/openroaming/} consortium hub. However, adherence to certain requirements is necessary for joining the hub.

OpenRoaming is a initiative led by the Wireless Broadband Alliance (WBA) to simplify Wi-Fi connectivity and roaming on a worldwide scale. By creating a global network of trust, OpenRoaming enables a seamless and secure connection experience, eliminating the need for repetitive authentication in different networks, making wireless connectivity more convenient and efficient for its users. One of the consortium's pillars is Cybersecurity, so it ensures that roaming between thousands of networks is done securely and encrypted.

The approach employed by these initiatives aims to enable users of mobile devices to automatically connect to Wi-Fi networks securely regardless of their location. In addition to RNP's interest in joining the OpenRoaming consortium, the integration of current federated infrastructure and 5G networks opens up possibilities for increased coverage area. This is aligned with the interests of the aforementioned consortium\footnote{https://wballiance.com/openroaming/faq/\#12ff1f427cfb6f8ab}.

The 3rd Generation Partnership Project (3GPP) is a global collaborative organization responsible for developing telecommunication standards, especially in the context of mobile networks. The architecture of the 5G networks is defined through publicly available documents called Releases. Releases are detailed technical documents that guide the implementation of telecommunication technologies in production environments.

Within this context, the analysis of supported authentication methods along with associated attributes, as well as the assessment of potential security enhancements in the integration and convergence of non-3GPP technologies is of utmost importance.

The research's key achievements can be outlined as:

\begin{enumerate}
    \item The investigation of literature surrounding the theme of device authentication in 5G networks, considering identity management (IdM) aspects.
    \item Construction and documentation of a testing environment that is used for validating some open-source 5G network tools and projects.
    \item Architectural study, based on the latest specifications, that possibly enables the convergence between WiFi and 5G in the Eduroam federation.
    \item Exploration of identity requirements in the context of the Eduroam federation to support the authentication of 5G mobile devices.
\end{enumerate}

The remainder of this paper is structured as follows:
Section~\ref{sec:background} offers a thorough explanation of key concepts vital for understanding the investigated subject. Section \ref{sec:results} gathers the results of both theoretical and practical stages of this work, while also presenting discussions regarding some of these results. Finally, Section \ref{sec:conclusion} outlines the main conclusions and delves into potential paths for future investigation.

\section{Background}
\label{sec:background}

It is necessary to explore concepts related to the Extensible Authentication Protocol (EAP) methods (covered in Section \ref{subsec:eap-fmk}) and the 5G Core (5GC) (Section \ref{subsec:5gc-analysis}), examine some of the 3GPP's Releases (whose main points are highlighted in Section \ref{subsec:3gpp-rels}), and also explore the EAP-AKA’ protocol (detailed in Section \ref{subsec:eap-aka-prime-details}), as these are closely related to central elements of this research.

\subsection{EAP Framework}
\label{subsec:eap-fmk}

\begin{figure}
    \centering
    \includegraphics[width=1.0\linewidth]{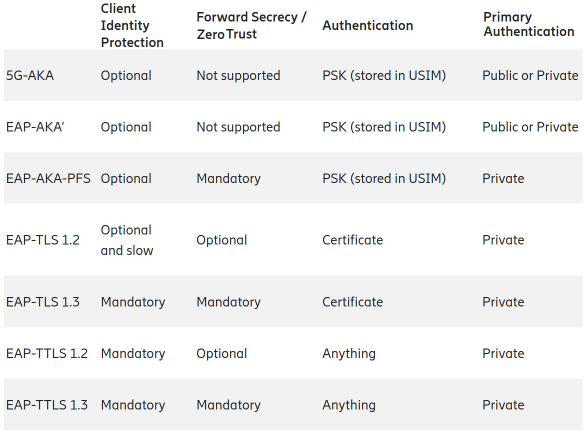}
    \caption{Main aspects of the available EAP methods for 5G. Source: \cite{eap}.}
    \label{fig:eap-methods}
\end{figure}

EAP is an authentication framework defined by RFC 5247 \cite{rfc5247}. It currently covers more than 40 authentication methods and is widely used in computer networks, being frequently employed in, for example, WiFi networks (where it is implemented within the IEEE 802.1X protocol). In a production environment, a database or directory containing attributes related to user or device access credentials is typically used. Both authentication and authorization processes revolve around these attributes. When it comes to 5G networks, the main authentication methods available are those listed in \figurename~\ref{fig:eap-methods}, which also presents the key characteristics of each. It is important to note that, according to specifications, only the 5G-AKA and EAP-AKA’ methods are valid for use in public 5G networks, while the other EAP methods remain as alternatives for private 5G networks.

In one hand, 5G-AKA is a key agreement method defined by 3GPP itself, which is not formally specified by a Request for Comments (RFC), and it is also not part of the EAP framework. On the other hand, the EAP-AKA’ is an authentication method defined by RFC 9048 \cite{rfc9048}, which, in the context of 5G, is employed to enable the generation and validation of an anchor key. This key serves as the basis for deriving section keys between the User Equipment (UE) and the gNodeB (gNB). It is noteworthy that the EAP-AKA’ protocol can also be used for device authentication in WiFi networks.

Both methods are improvements of earlier protocols, such as EPS-AKA’ in the case of 5G-AKA and EAP-AKA in the case of EAP-AKA’. According to the results presented in \cite{eapaka}, despite both having similar features in terms of performance and privacy, when directly compared, there are still differences regarding security aspects. 
EAP-AKA’ adds some identifiers as Serving Network ID (SNID), Subscription Permanent Identifier (SUPI), and Subscription Concealed Identifier (SUCI) in the key generation process. It also makes improvements to the inputs used by the Key Derivation Functions (KDFs), adds three EAP Type section IDs, a Random Challenge (RAND), and an Authentication Token (AUTN).
Extra details regarding the protocol's operation will be presented in Section \ref{subsec:eap-aka-prime-details}.

\subsection{3GPP Releases}
\label{subsec:3gpp-rels}

Considering the context of this work, Releases 15, 16, and 17 can be highlighted as the most relevant. Due to the considerable scope of these documents, only their main points are explicitly detailed in the following paragraphs.

From Release 15 onwards, the architecture of 5G includes the group of components that make up the 5GC. These components are called Network Functions (NFs) in this context. They are network elements or software instances that are specified and/or adopted by 3GPP and have clearly defined functional behavior and interfaces. The main goals of 5GC are to be flexible (allowing virtualization) and to manage network resources to connect the UE to the Data Network (DN) (which is usually the internet) and/or other resources available in the network. During the early versions of this Release, it was necessary to transition from the previous generation (4G) to the recently launched one; therefore, the specified network architecture was called Non-Stand-alone Architecture (NSA).

In Release 16, non-3GPP technologies were included in the 5G architecture by 3GPP. The NF called Non-3GPP Inter Working Function (N3IWF) is the main component of the 5G architecture that provides the integration functionality of non-3GPP networks and devices. The connection of UEs can be made through other wireless access technologies, enabling support for legacy devices known as Non-5G Capable (N5GC). In general, the access of non-3GPP UEs occurs in such a way that they forward their traffic to a trusted gateway within the infrastructure of the Mobile Network Operator (MNO) using the N3IWF to establish the N2 and N3 communication interfaces. This Release introduced a new setup called Stand-alone Architecture (SA). In this new configuration, the network architecture is set up to have only 5G networks end-to-end. Thus, the improvements brought by the new generation could be better explored, confirming the advancements of 5G compared to previous generations.

Finally, in Release 17, the component called Data Network Authentication, Authorization, and Accounting (DN-AAA) was included. In conjunction with the elements of 5GC represented in \figurename~\ref{fig:dn-aaa-arch}, the DN-AAA can participate in the authentication of UEs in the 5G infrastructure via the N6 interface. Another main highlight is the specification of EAP methods for 5G and the possibility of including a RADIUS-type server in the network architecture (through DN-AAA NF). This server is currently in use in the device authentication processes by Eduroam in Brazil.

\subsection{Analysis of 5G Core Projects}
\label{subsec:5gc-analysis}

\begin{figure}
    \centering
    \includegraphics[width=1.0\linewidth]{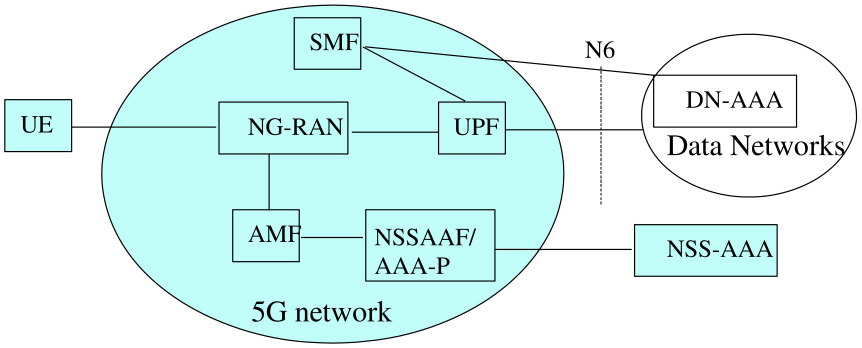}
    \caption{Reference architecture for authentication-related network functions according to the Release 17. Source: \cite{3gpp1}.}
    \label{fig:dn-aaa-arch}
\end{figure}

The 3GPP specification for the 5GC \cite{3gpp3} features the presence of several NFs. Among these, this work focused on the use of those directly involved in authentication procedures. Considering the specification from Release 17 \cite{3gpp1} and the theoretical background of this work, it is possible to highlight \figurename~\ref{fig:dn-aaa-arch}, which presents the 5GC NFs involved in UE authentication and authorization processes. More details about this architecture and its operation are addressed in Section \ref{subsec:eap-aka-prime-details} and Section \ref{sec:results}.

The technical report \cite{br1} of the Brasil 6G project contains a detailed analysis of open-source projects for the 5GC component of the 5G network infrastructure. Among the alternatives analysed, two stand out: Open5GS and free5gc.

The Open5GS project has a well-organized source code, mostly written in the C language, and a good level of documentation. However, given the possibility of integrating a gNB created from another research project \cite{br2} and the focus on integrating the solution with the existing WiFi network infrastructure of the Eduroam federation, the presence of the N3IWF component was mandatory. At the time of choosing the 5GC solution for this work, this component was not implemented in the Open5GS project.

Contrasting the evaluation results of Open5GS, the free5gc project already had the N3IWF implemented. The project, which has most of its source code written in Go, benefits from a high level of modularity of its components. It also benefits from a support documentation\footnote{https://free5gc.org/guide/}, which includes tutorials to assist its users. As pointed out in the report, free5gc has gained considerable acceptance within the scientific community and has a more permissive software license. This would allow, for example, the use of a customized version of the software without the need to publicly release the modified source code.

In addition to enabling communication between the 5GC and experimental devices (not approved by 3GPP), the N3IWF component also facilitates communication with various other types of non-3GPP Radio Access Networks (RANs), such as WiFi. This allows supporting the connection of network interfaces for N5GC devices.

\subsection{EAP-AKA' Protocol Detailed}
\label{subsec:eap-aka-prime-details}

The study \cite{eapaka} presents a considerably detailed formal analysis of the EAP-AKA’ protocol, exploring the mathematical proofs behind this protocol and dissecting its internal operations. As part of the verification performed, the authors instantiate a 5G test environment based on the NS-3 simulator\footnote{https://www.nsnam.org/}. Subsequently, the EAP-AKA’ is implemented in an open-source validation software called ProVerif\footnote{https://bblanche.gitlabpages.inria.fr/proverif/}, and from this, an analysis and subsequent comparison of EAP-AKA’ with 5G-AKA follows. These are the protocols specified for usage in public 5G networks.

After reviewing the content presented in \cite{eapaka} and combining it with the slightly more concise approach brought in the survey conducted by the work \cite{5gauth}, it was possible to determine which steps were most important for understanding the EAP-AKA’ protocol in the context of 5G.

\begin{figure*}
    \centering
    \includegraphics[width=1.0\linewidth]{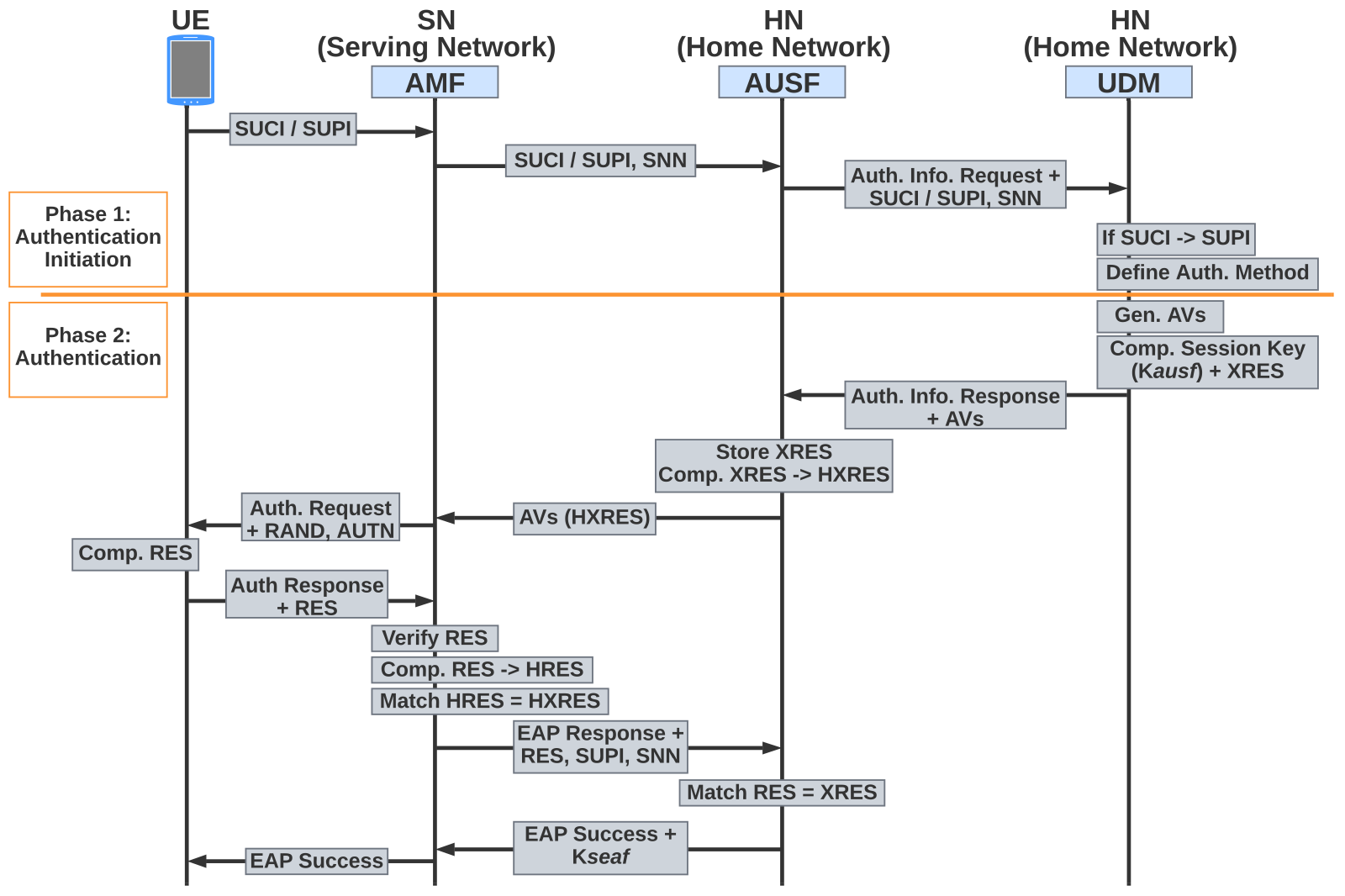}
    \caption{Main procedures of EAP-AKA' algorithm in 5G environments. Based on \cite{eapaka} and \cite{5gauth}.}
    \label{fig:aka-prime-diagram}
\end{figure*}

Thus, \figurename~\ref{fig:aka-prime-diagram} highlights the main operations performed by the EAP-AKA’ algorithm in the context of UE authentication in a 5G network.

Initially, the concepts of Home Network (HN) and Serving Network (SN) need to be explained. HN represents the network where the user is originally registered and SN is the network where the user is currently connected. According to the specification, the Unified Data Management (UDM) and Authentication Server Function (AUSF) elements should belong to the HN. While the Access and Mobility Management Function (AMF) is usually located in the SN as in the example depicted in \figurename~\ref{fig:aka-prime-diagram}. However it could also be placed in the HN since there is the possibility of the SN being UE's own HN.

Didactically, the authentication procedure can be divided in two phases: 1) Authentication Initiation; and 2) Authentication. As soon as an identity request is sent by the AMF or, depending on the configuration of the 5GC, by the Security Anchor Function (SEAF), the first phase of the process begins.

In the first step, the UE sends an identifier (which can be the SUCI or the SUPI) which is received by the AMF. The AMF then concatenates this identifier with the Serving Network Name (SNN) and sends it to the AUSF. The AUSF sends an authentication information request to the UDM along with the data received from the AMF. Upon receiving this request, the UDM extracts, if necessary, the SUPI from the SUCI (a procedure already provided by the specification \cite{3gpp4}) and concludes the first phase of the process by defining which authentication method will be used for that UE. As the selected algorithm influences the next steps and due to the context of this research, in this example, EAP-AKA’ was chosen.

Starting the second phase of the process, the UDM generates the Authentication Vectors (AVs) and also computes the anchor key ($K_{ausf}$) and the Expected Response (XRES). In the next step, the UDM sends an authentication information response and the AVs back to the AUSF. The XRES is stored in the AUSF for future use, and the Hashed Expected Response (HXRES) is computed from the received XRES. The AMF then receives the AVs (which now carry HXRES instead of XRES) and creates an authentication request for the UE, which is concatenated with RAND and AUTN. Consecutively, the UE computes the Response (RES), concatenates it with an authentication response, and sends it to the AMF. In this second phase, the AMF verifies the RES, computes the Hashed Response (HRES) from it, and compares HRES with the HXRES that had been received previously via AVs. If all these conditions are met, the AUSF receives an EAP response concatenated with RES, SUPI, and SNN from the AMF, comparing the received RES with the XRES stored earlier. Then, the AMF receives an EAP Success message along with an SEAF key ($K_{seaf}$) from the AUSF and then forwards the EAP Success message to the authenticating UE.

\section{Results and Discussion}
\label{sec:results}

To support the investigation of EAP authentication methods in 5G networks, some open-source tools were chosen to build a test environment with one of its main objectives being the validation of these tools.

\begin{figure}
    \centering
    \includegraphics[width=0.65\linewidth]{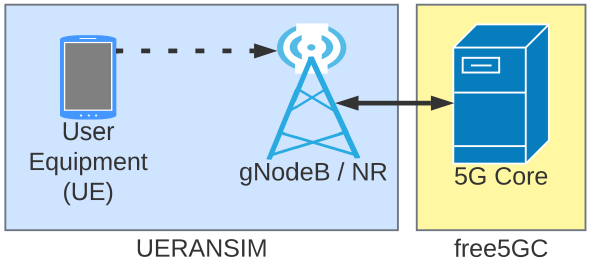}
    \caption{Testing environment for 5G networks.}
    \label{fig:test-env}
\end{figure}

\begin{figure}
    \centering
    \includegraphics[width=1.0\linewidth]{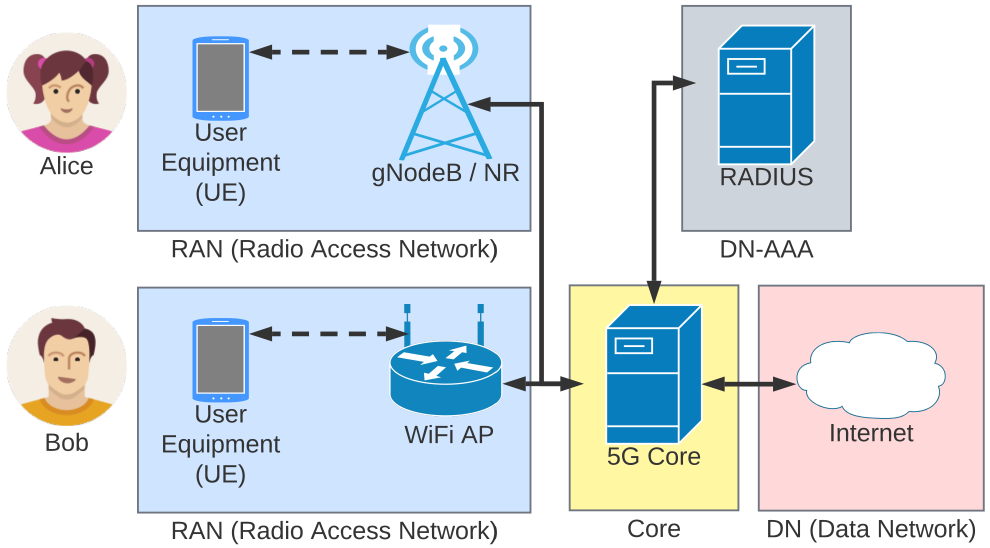}
    \caption{Convergence architecture between WiFi and 5G based on the specification \cite{3gpp1} and suitable to be applied in the Eduroam federation.
    }
    \label{fig:convergence-arch}
\end{figure}

Based on the results obtained from the literature analyses mentioned in Section \ref{sec:background}, the free5gc project was used as the 5GC element. One of the UE and RAN simulators recommended by the developers of the free5gc project is UERANSIM; therefore, \figurename~\ref{fig:test-env} presents the test environment. From the environment represented in \figurename~\ref{fig:convergence-arch} and the knowledge accumulated during the theoretical phase of the project, some tests were conducted, and the exploration of a convergence architecture was prospectively carried out.

\figurename~\ref{fig:convergence-arch} presents an overview of the network architecture that would enable convergence between WiFi and 5G technologies in the context of the Eduroam federation. The diagram illustrates the connection of two different users (Alice and Bob) to the internet (expressed in the 5G architecture by the NF DN). Alice connects via the 5G network, while Bob connects via the WiFi network; nevertheless, both wireless network devices in the RAN communicate with the 5GC, and it authenticates both users on a RADIUS server through the NF DN-AAA. It is essential to note that, despite the didactic separation of Alice and Bob's identities, in the context of a handover operation, the identity could be the same. In this case, the RAN technology would change over time (i.e., the user connected via WiFi switches to connect via 5G, or vice versa). The handover operation is specified for 5G in \cite{3gpp5} through the Access Traffic Steering, Switching, and Splitting (ATSSS) technique, which, in addition to the transparent change of access technology, would also enable traffic offloading through the simultaneous use of multiple RANs.

In spite of that, there are still open questions regarding the practical implementation of this architecture, one of them being the integration of credentials from different environments (mobile networks and WiFi networks). Credentials in a mobile network are generally based on the International Mobile Subscriber Identity (IMSI), but an Eduroam identity is based on the junction of an identifier with a realm (the latter being the authentication server's domain address).

To illustrate this issue, it is possible to see in Listing \ref{src:auth} an example of an authentication attempt by a UE (mobile network) on the Eduroam network (WiFi).

\begin{lstlisting}[caption={Authentication attempt within Eduroam federation.},captionpos=b,label={src:auth}]
Sun May 22 00:03:13 2022: Access-Reject 
for user 6724313930974708@wlan.mnc031.
mcc724.3gppnetwork.org stationid 
84-37-D5-B3-49-F1 from _self_ 
(Misconfigured client: Unsupported 3G 
EAP-SIM client! Rejected by <TLD>.) to 
eduroam01.ufpe.br (150.161.50.4)
\end{lstlisting}

It is feasible to say that an investigation around the possibility of associating identifiers is needed, still it is also necessary to consider user privacy. In this regard, the investigation can be directed towards the use of Temporary Mobile Subscriber Identity (TMSI). Nevertheless, the definitive solution to handle this situation in practice is still open, yet it may involve some address translation and/or encapsulation technique.

In terms of identity attributes, as the main requirements for adopting the convergence architecture presented in \figurename~\ref{fig:convergence-arch}, it would be necessary to add a device identifier (which, according to specification \cite{3gpp2}, should be the IMSI or TMSI). Other attributes that will need to be included in the system, according to the key generation process addressed and represented in \figurename~\ref{fig:aka-prime-diagram}, are the triplets (or 5-tuples), which are a requirement for generating AVs. These may be stored in a database or in the Home Location Register (HLR) (an element that belongs to the infrastructure of a telecommunication's operator) and not directly in the user's profile.

As part of the results of this work, additional documentation was created to assist in reproducing the results, and packet capture files were generated to aid in future static analyses of the instantiated test environment. All these artifacts are available online\footnote{https://github.com/oliveiraleo/RNP-PGId-2023} on a GitHub repository. During the execution of the mentioned activities, and based on the findings and experiences gained in the process, various contributions\footnote{https://github.com/free5gc/free5gc.github.io/commits?author=oliveiraleo} were also made to the official documentation of the free5gc project.

\section{Conclusion}
\label{sec:conclusion}

It is noticeable that, in facing the disruptive characteristics of 5G, such as the paradigm shift and the new network architecture, the relevance of acquiring the background information intensifies. A deeper understanding provides greater clarity and security in proposed solutions, even with the intrinsic complexity of the implementation phase.

The project involved an investigation into literature, standards, and specification documentation surrounding device authentication in 5G networks. In the theoretical phase, EAP methods applied to 5G networks were studied. During the practical phase, open-source tools resulting from the theoretical investigation were validated, meeting nearly all requirements. Result's reproducibility were facilitated through documentation of the test environment, which is available for future reuse in other projects. The acquired knowledge was also successfully applied to solutions suitable for future setup in the Eduroam federation.

As part of future work, there is a plan to create a RADIUS experimental environment for testing the EAP-AKA' protocol. Additionally, the feasibility of execution of roaming between different networks or institutions within the federation should be verified. Exploring the usage of other EAP methods for private 5G networks is also under consideration. Furthermore, testing with a real UE in a 5G environment based on the constructed setup is planned. Finally, the goal is to culminate in building an experimental environment containing a WiFi network, a 5G network, and a RADIUS authentication server to investigate the possibility of native authentication for devices in mobile and WiFi networks simultaneously, along with the practical exploration of some of the mentioned techniques.

\section*{Acknowledgment}

We would like to acknowledge the Networks and Distributed Systems Laboratory (NetLab) of the Federal University of Juiz de Fora (UFJF) for providing the network infrastructure used to carry on the experiments of this research.

\end{document}